\documentclass{article}
\usepackage{graphicx} 
\usepackage{amsmath}
\usepackage{amssymb}
\usepackage{physics}
\usepackage{cite}
\usepackage{geometry}
\usepackage{mathrsfs}
\usepackage{mathtools,leftindex,tensor,mhchem}
\usepackage{authblk}
\usepackage{breqn}
\usepackage{subcaption}
\usepackage{float}
\usepackage[pagebackref=false, colorlinks=true]{hyperref}
\hypersetup{
linkcolor=blue,     
citecolor=blue,     
urlcolor=blue}

\title{The Ultra Slow-Roll Phase Of Warm Inflation In Braneworld Cosmology }
\author{Aarav Shah \thanks{Electronic address:- shahaarav103@zohomail.in} }
\affil{International Center For Space and Cosmology, Ahmedabad University, Ahmedabad 380009,Gujarat,India }

\date{\today}
\begin{document}

\maketitle

\begin{abstract}
Slow-roll of the inflaton (inflationary field) defines the standard dynamics of the inflationary epoch. However,
the inflaton deviates from slow-roll when it encounters an extremely flat region in the
inflationary potential, and enters a phase dubbed  Ultra Slow Roll (USR). In previous studies, there have been various theories which modify the theory of general relativity, all of them having different motivations based on different paradigms. Among these, braneworld gravity, motivated from string theory; is one of the most prominent theories as it provides a geometrical explanation for the weakness of gravity.  In this article, we explore two possible braneworld background theories, the Randall-Sundrum (RS-II) model and the Dvali-Gabadadze-Porratai (DGP) model, and then realize an USR phase in a particularly interesting inflationary scenario, called warm inflation. In the warm inflationary scenario, a thermal radiation bath coexists with the inflationary energy density as an effect of the dissipative dynamics. We then derive inflationary slow roll parameters and  the primordial power spectrum of scalar curvature perturbations in such a setup. We then  numerically investigate the evolution of the inflaton  and the primordial power spectrum of scalar curvature perturbations. Our analysis shows that the braneworld contributions become progressively suppressed as the USR conditions are made more stringent, indicating that the USR phase effectively diminishes brane-induced corrections to standard inflationary dynamics.
\end{abstract}

\section{Introduction}
The cosmological principle of large-scale homogeneity and isotropy remains one of the fundamental assumptions which shape up modern cosmology\cite{Baumann:2022cosmology,Mukhanov:2005fem,Weinberg:2008cosmology,Dodelson:2003modern,Liddle:2000introduction,Kolb:1990early}. However, in order to make the large-scale homogeneity of the universe consistent with the principles of relativity and causality, the theory of an accelerated expansion of the universe (i.e inflation \cite{guth1981,Linde:1982new,Albrecht:1982eternal,Lyth:2009inflation}) arose in the 1980s. Since then, the inflationary paradigm has been of great interest in the scientific community, and subsequently, the field has seen various developments Numerous models of inflation have been proposed; each having their own advantages and disadvantages.
\\
\\
Several studies point out that during inflation, the potential could transiently become extremely flat, giving rise to a short-lived ultra slow-roll phase\cite{Villanueva_Domingo_2021,guth1981,kazanas1980,albrecht1982,linde1982,sato1981b,sato1981a}. The flatness of the potential induces an exponential dependence of the power spectrum for scalar curvature perturbations with the number of e-folds and thus leads to an enhanced formation of primordial black holes (PBH) which are viable candidates for dark matter \cite{young2024computingabundanceprimordialblack,pi2025primordialblackholeformation,Motohashi_2017,Carr_2016}. Such a phase of inflation is called the Ultra Slow Roll (USR) phase. 
\\
\\
In the standard model of inflation (namely, the cold, slow roll inflation), we neglect dissipative effects of the inflaton into the radiation bath of the early universe. This leads to various problems in transitioning from the inflationary phase to standard cosmology where the universe doesn't undergo accelerated expansion. Previous studies have proposed various theories like reheating of the radiation bath so that the universe proceeds with the hot big bang phase as expected from standard cosmology. As an alternative Berera \cite{Berera_1995}, proposed the warm inflation model where the inflaton  dissipates energy to the radiation bath in such a way that a constant radiation density is maintained \cite{Kamali_2023,Berera_2023,Trivedi:2020ljd}. Furthermore, in a recent work \cite{Biswas_2024}, the possibility of realizing USR dynamics to warm inflation was considered in a general relativistic background. 
\\
\\
A further layer of interest emerges when one considers braneworld cosmology \cite{Maartens_2010,Langlois_2002,Brax_2004}, particularly the Randall–Sundrum type II (RS-II) and Dvali-Gabadadze-Porratai (DGP) scenario \cite{Deffayet2001,Sahni2003,Lue2006,Maartens2006}, where our 4-dimensional universe is a brane embedded in a higher-dimensional bulk. The RS-II framework modifies the Friedmann equation at high energies by introducing quadratic energy-density corrections, which significantly alter the dynamics of the early Universe.  A few models of inflation have been studied before in the RS-II background and the DGP background by works like \cite{Herrera_2011,Cid_2007,Panotopoulos_2007,Trivedi:2021ivk} and \cite {BouhmadiLopez2011,Hossain2016} respectively. The general conclusion of these works is that the RS-II model has significant impacts on early universe cosmology and not on late universe cosmology; while, the DGP model has significant impacts on late universe cosmology and not on early universe cosmology. This is because, RS-II induced corrections are only prevalent at the high energy regimes (i.e the early universe) while DGP induced corrections are only prevalent during low energy regimes (i.e the late universe).
\\
\\
Continuing from this line of study, our work aims to consider the USR phase of warm inflation in a RS-II and DGP braneworld cosmology background. We analyze the additional effects on the evolution of the inflaton and the primordial power spectrum induced due to the braneworld setup in this scenario. This work is structured as follows. In Sec. \ref{Section 2}, we give a short overview of various aspects of our analysis which includes the RS-II braneworld cosmology, the USR phase in cold inflation and the warm inflation model. In  Sec. \ref{Section 3} and \ref{Section 4} we present the dynamics of the inflaton during the USR phase of warm inflation in the RS-II braneworld setup and the DGP model respectively. We then derive analytical expressions for the slow-roll parameters, the primordial power spectrum for curvature perturbations and the number of e-folds between two values of the inflaton in such a model. Sec.  \ref{Section 5} then illustrates numerical results and plots for the evolution of the inflaton and the primordial power spectrum of scalar curvature perturbations for representative choices of the potential (namely linear and cubic potentials) and extreme cases of various parameters like the dissipation coefficient, the brane tension and dark radiation. We conclude our work in Sec. \ref{Conclusion} with a summary.

\section{Preliminaries - An Overview Of Braneworld Gravity, USR Dynamics And Warm Inflation} \label{Section 2}

In string theoretic frameworks, the observable universe is often modeled as a $4$ dimensional ($1+3$) brane embedded in a $5$ dimensional ($1+4$) bulk space-time, giving rise to the framework known as  braneworld gravity \cite{Germani_2002,Maartens_2010,Langlois_2002,Brax_2004}. 
Studying braneworld gravity in detail requires us to briefly review notions of the extrinsic curvature and the Gauss-Codazzi equations and subsequently make use of them in the context of braneworld gravity. \footnote{We refer the unfamiliar reader to standard texts Carmo\cite{doCarmo2016}, O'Niel \cite{ONeill1983}, or Poisson \cite{Poisson2004} for a more detailed treatment on these concepts of differential geometry.} 
\\
\\
The extrinsic curvature $K_{\mu\nu}$ characterizes how the brane is curved within the bulk and it is defined as
\begin{equation}    K_{\mu\nu}=h^{\alpha}_{\mu}h^{\beta}_{\nu}\nabla_{\alpha}n_{\beta},
\end{equation}
where $n_{\beta}$ is the unit normal to the brane, $h_{\mu\nu}$ is the induced metric on the brane and $\nabla_a$ is the covariant derivative in the bulk. Intuitively, $K_{\mu\nu}$ measures how neighboring points on the brane bend relative to each other in the extra dimension, and it enters the effective $4D$ Einstein equations through the Gauss-Codazzi relations, contributing extra terms that distinguish braneworld from standard general relativity.
\\
\\
The Gauss equation gives us the $4D$ curvature tensor in terms of the projection of the $5D$ curvature,
\begin{equation}\label{Gauss}
R_{ABCD}
= \prescript{(5)}{}{R}_{EFGH}\, g^{E}{}_{A} g^{F}{}_{B} g^{G}{}_{C} g^{H}{}_{D}
  + 2\, K_{A[C} K_{D]B}.
\end{equation}
where the square brackets denote anti-symmetrization. On the other hand, the Codazzi equation determines the change of $K_{AB}$ along the brane via 
\begin{equation}
    \nabla_{B}K^{B}_{A}-\nabla_{A}K=\prescript{(5)}{}R_{BC}g_{A}^{B}n^{C},
\end{equation}
where $K=K^{A}_{A}$ is the trace of the extrinsic curvature. Now, let us return to the Einstein field equation and let $\Lambda_5$ denote the cosmological constant in the braneworld set-up. The $5D$ field equations determine the $5D$ curvature tensor where in the bulk, they are given as
\begin{equation} \label{5DEFT}
    \prescript{(5)}{}G_{AB}=-\Lambda_5\prescript{(5)}{}g_{AB}+\kappa_5^2\prescript{(5)}{}T_{AB},
\end{equation}
where we note that $\prescript{(5)}{}T_{AB}$ represents the $5D$ energy-momentum of the gravitational sector and $\kappa_5=8\pi G$ in natural units ($c=1$), $G$ being the gravitational constant. From eqns. \eqref{Gauss} and \eqref{5DEFT}, we can write
\begin{align}
    G_{\mu\nu}=-\frac{1}{2}\Lambda_5g_{\mu\nu}+\frac{2}{3}\kappa_5^2\left[\prescript{(5)}{}T_{AB}g_{\mu}^{A}g_{\nu}^{B}+\left(\prescript{(5)}{}T_{AB}g_{\mu}^{A}g_{\nu}^{B}-\frac{1}{4}\prescript{(5)}{}T\right)g_{\mu\nu}\right]\\+KK_{\mu\nu}-K_{\mu}^{\alpha}K_{\alpha\nu}+\frac{1}{2}\left[K^{\alpha\beta}K_{\alpha\beta}-K^2\right]g_{\mu\nu}-E_{\mu\nu},
\end{align}
where $\prescript{(5)}{}T=\prescript{(5)}{}T^{A}_{A}$ is the trace of the energy-momentum tensor and where \footnote{The interested reader can refer to standard texts like \cite{Maartens_2010,Langlois_2002} about braneworld gravity.}
\begin{equation}
    E_{\mu\nu}=\prescript{(5)}{}C_{ABCD}n^Cn^Dg_{\mu}^Ag_{\nu}^B
\end{equation}
is the projection of the bulk Weyl tensor orthogonal to $n^A$.
\\
\\
We now introduce a Brane-tension $\lambda$  as an effective cosmological constant for the $1+3+d$-dimensional spacetime. We can write
\begin{equation}
    \prescript{(5)}{}T_{\mu\nu}=T_{\mu\nu}-\lambda g_{\mu\nu}.
\end{equation}
When then confine our interests to a Friedmann brane, and we enter the realm of cosmology. The Einstein field equations change in braneworld cosmology and one can write the following modified Friedmann system
\begin{equation} \label{f1}
    H^2=\frac{\rho}{3M_{Pl}^2} \left({ 1+\frac{\rho}{2\lambda}}\right)+\frac{\rho_{E_0}}{3M_{Pl}^2}\left(\frac{a_0^4}{a^4}\right)+\frac{\Lambda_5}{3}-\frac{K}{a^2}
\end{equation}
where $\rho_{E}$ can be interpreted as the "dark radiation" term coming from the bulk and can be thought of the density induced due to the projection $E_{\mu\nu}$ of the  bulk Weyl tensor orthogonal to $n^A$ (see \cite{Maartens_2010,Langlois_2002} for more information). 
Differentiating eqn. \eqref{f1} with respect to time yields
\begin{equation} \label{f2}
    \dot{H}=-\frac{1}{2M_{Pl}^2}\left(\rho+p \right)\left(1+\frac{\rho}{\lambda}\right)-\frac{2\rho_{E_0}}{3M_{Pl}^2}\left(\frac{a_0^4}{a^4}\right)+\frac{K}{a^2}.
\end{equation}
\\
\\
But eqn. \eqref{f1} is only really valid for one particular type of Braneworld. On a more fundamental level, in Braneworld gravity the gravitational dynamics depend crucially on the structure of the total action where one specifies how the brane is embedded in the higher-dimensional bulk and how gravity behaves across this interface. Two of the most prominent realizations of this framework are the Randall–Sundrum type II (RS-II) model \cite{rs1,rs2,Gogberashvili} and the Dvali–Gabadadze–Porrati (DGP) model \cite{dgp}. While both introduce an extra spatial dimension, their treatment of gravity and localization mechanisms differ fundamentally. In the RS-II scenario, our universe is a 3-brane embedded in a 5-dimensional Anti–de Sitter (AdS) bulk with a negative cosmological constant and the full action includes contributions from both the 5-dimensional bulk and the 4-dimensional brane surface
\begin{equation}\label{RS-IIaction}
S_{\text{RS-II}}=\int \sqrt{-g_5}\left(\frac{1}{2\kappa_5^2}R_5-\Lambda_5\right) d^5x +\int \sqrt{-g}\left(-\lambda+\mathcal{L}_{\text{matter}}\right) d^4x
\end{equation}
where $R_5$ is the 5-dimensional Ricci scalar, $\Lambda_5<0$ is the bulk cosmological constant and $\lambda$ is the brane tension. The matter fields are confined to the brane, and $\kappa_5^2=8\pi G_5$ defines the 5D gravitational coupling. The negative bulk cosmological constant balances the positive brane tension, and this ends up ensuring us that the effective 4D cosmological constant vanishes and gravity becomes localized near the brane through the exponential warping of the bulk metric. This localization then arises naturally from the warped metric
\begin{equation}
ds^2=e^{-2k|y|}g_{\mu\nu}(x)dx^\mu dx^\nu+dy^2
\end{equation}
where $k=\sqrt{-\Lambda_5/6}$ is the curvature scale of the AdS bulk.
\\
\\
In contrast, the DGP model modifies the brane action by introducing an induced 4-dimensional Einstein–Hilbert term directly on the brane, which ends up altering the gravitational dynamics at large distances and the DGP action is given as
\begin{equation}\label{DGPaction}
S_{\text{DGP}}=\int d^5x\sqrt{-g_5}\left(\frac{1}{2\kappa_5^2}R_5\right)+\int d^4x\sqrt{-g}\left(\frac{1}{2\kappa_4^2}R+\mathcal{L}_{\text{matter}}\right)
\end{equation}
The ratio of the two couplings defines a characteristic crossover scale which is given as
\begin{equation}
r_c=\frac{\kappa_5^2}{2\kappa_4^2}=\frac{M_4^2}{2M_5^3}
\end{equation}
beyond which gravity transitions from 4-dimensional to 5-dimensional behavior. Now one sees that at scales $r\ll r_c$, gravity behaves as in standard $4D$ general relativity whereas at cosmological scales $r\gg r_c$, gravity “leaks” into the extra dimension, leading to late-time cosmic acceleration without invoking a cosmological constant.
\\
\\
The effective Friedmann equations for the DGP cosmology exhibit this departure as they show a square root structure characteristic of brane-induced gravity
\begin{equation}\label{DGPEFE}
H^2=\left(\sqrt{\frac{\rho}{3M_{\text{Pl}}^2}+\frac{1}{4r_c^2}}+\frac{\epsilon}{2r_c}\right)^2
\end{equation}
where $\epsilon=\pm1$ labels the two possible branches: the self-accelerating branch ($\epsilon=+1$) and the normal branch ($\epsilon=-1$). The RS-II model modifies gravity through high-energy corrections proportional to $\rho^2$ thus leading to eqn. \eqref{f1}, while the DGP model alters it through nonlocal modifications arising from brane-induced curvature. 
\\
\\
We now turn our attention to inflationary cosmology. To understand how warm inflation behaves when the potential becomes extremely flat (USR phase of inflation) in a braneworld cosmology, we shall first understand the  USR phase \cite{Dimopoulos_2017,Martin_2013,Kinney_2005,Di_Marco_2024}  in an $FLRW$ space-time in the corresponding cold inflation case \footnote{Cold inflation really refers to just the conventional inflationary scenario where the scalar field does not decay during the inflationary phase.} and in cold inflation models \cite{guth1981,kazanas1980,albrecht1982,linde1982,sato1981b,sato1981a} the inflaton $\phi$ evolves as
\begin{equation}
    \ddot{\phi}+3H\dot{\phi}+V_{,\phi}=0
\end{equation}
where $V_{,\phi}$ is the usual derivative of the scalar potential with the field and the expansion of the universe induces a friction term $3\dot{H}\phi$ in the equation of motion for the scalar inflaton $\phi$. The energy density and pressure  corresponding to the inflaton $\phi$ are given by $\rho_{\phi}=\frac{\dot{\phi}^2}{2}+V(\phi)$ and $P_{\phi}=\frac{\dot{\phi}^2}{2}-V(\phi)$.  The Friedmann equation for $H^2$ now takes the form
\begin{equation}
    3M_{Pl}^2H^2=\frac{\dot{\phi}^2}{2}+ V(\phi),
\end{equation}
where $M_{Pl}$ is the reduced Planck mass. We now define the slow roll parameters as
 \begin{equation}
     \epsilon_1=\frac{-\dot{H}}{H^2}
 \end{equation}
 and 
 \begin{equation}
     \epsilon_2=\frac{\dot{\epsilon_1}}{\epsilon_1H}
 \end{equation}
 Note that, we have a successful theory of inflation if and only if $|\epsilon_1|<<1$. In standard cold inflation, there is a further condition that $|\epsilon_2|<<1$ as we shall see below, this condition is not valid in the USR phase of inflation; in fact, $|\epsilon_2|\simeq 6$.
 \\
 \\
The Friedmann equations imply that
 \begin{equation} \label{Extremely flat}
     \epsilon_2=-6-\frac{2V_{,\phi}}{H\dot{\phi}} +2\epsilon=\frac{2\ddot{\phi}}{H\dot{\phi}}.
 \end{equation}
 If the potential becomes extremely flat then, by eqn. \eqref{Extremely flat}, we have $\epsilon_2\simeq -6+2\epsilon_1$. Hence, $|\epsilon_2|\simeq 6$ contrary to the standard cold inflation where $|\epsilon_2|<<1$ as commented above.
 However, one can prove that $|\epsilon_1|<<1$ still holds true and thus, the USR phase  is a valid inflationary phase.
 We now  comment briefly on the curvature perturbation during USR inflation. The power spectrum of scalar curvature perturbation in general reads
 \begin{equation}
     P=\left(\frac{H^2}{2\pi\dot{\phi}}\right)^2
 \end{equation}
 which using $-\dot{\phi}^2=2M_{Pl}\dot{H}$ can be written as
 \begin{equation}
     P=\frac{H^2}{8\pi^2M_{Pl}^2\epsilon_1}
 \end{equation}
 During USR, $\ddot{\phi}+3H\dot{\phi}=0$ and hence, $\dot{\phi}\propto a^{-3}$ and $\epsilon_1\propto a^{-6}$ and $P\propto a^6=e^{6\Delta N}$ where $N$ is the number of $e-$folds and hence, the curvature perturbation grows exponentially during the USR phase-leading to the formation of primordial black holes (see \cite{PhysRevD.105.083525}  for more information). With this, we shall conclude our review for USR dynamics and shall subsequently now turn our attention towards the warm inflation model.
\\
\\
 Warm inflation \cite{BASTERO_GIL_2009} has been widely used in $GR$, with several recent works reviewing its dynamics, the weak vs. strong dissipative regimes, and observational constraints \cite{2023Univ....9..124K}.
 During warm inflation \cite{Berera_1995,Kamali_2023,Berera_2023}, the inflaton  $\phi$, dissipates its energy to a radiation bath, maintaining a non-negligible radiation energy density $\rho_r$ throughout. This feature distinguishes warm inflation from the standard cold inflation scenario. Therefore, the equation of motion of the inflaton  also differs from that of the cold inflation scenario \footnote{Some works\cite{Motaharfar_2019,Das_2020,Kamali_2020} have even concluded that the new equation of motion, which arises due to the dissipative mechanism of warm inflation,  could possibly evade swampland conjectures \cite{Raveri:2018ddi,Hamaguchi:2018vtv,Fukuda:2018haz,Trivedi:2020wxf}. Furthermore, USR dynamics could also evade swampland conjectures \cite{Dimopoulos_2017,GargKrishnan2019BoundsSlowRollSwampland}.}. The equations governing the dynamics of the inflaton $\phi$, and the radiation bath, $\rho_r$, in warm inflation can be written as 
 \begin{equation}
     \ddot{\phi} + 3H\dot{\phi} +V_{,\phi}=-\Gamma (\phi,T)\dot{\phi}
 \end{equation}
 \begin{equation}\label{Raddis}
     \dot{\rho}_r+4H\rho_r=\Gamma(\phi,T)\dot{\phi}^2.
 \end{equation}
Here $\Gamma$  \cite{Benetti_2017} is the additional dissipative term induced by warm inflation which can depend on the amplitude of the field, $\phi$, as well as the temperature of the radiation bath $T$. Studies about warm inflation point out that 
\begin{equation}
    \Gamma =C_{\Gamma}T^p\phi^cM^{1-p-c}.
\end{equation}
 Here $C_{\Gamma}$ is a dimensionless constant carrying the signatures of the microscopic model used to derive the dissipative coefficient.  Note that studies show that warm inflation can occur at a finite temperature  $T>H$\cite{Bastero_Gil_2016}.
 \\
 \\
 We now define the dimensionless parameter $Q$ as the ratio of the two friction terms appearing in the equation of motion for the evolution of the inflaton $\phi$, one is due to dissipation while the other is due to expansion \cite{Bastero_Gil_2013}
 \begin{equation}\label{diss}
     Q=\frac{\Gamma}{3H}.
 \end{equation}
The equation of motion for the inflation $\phi$ now takes the simplified form
 \begin{equation}
     \ddot{\phi} +3H\dot{\phi}(1+Q)+V_{,\phi}=0.
 \end{equation}
 \begin{itemize}
     \item \textbf{Weak Dissipation}:- If the Hubble friction dominates over the friction due to dissipation; $Q<<1$ 
     \item \textbf{Strong Dissipation}:- If the Friction due to dissipation dominates over the Hubble friction then, $Q>>1$.
 \end{itemize}
 In the USR regime, the equation of motion for $\phi$ reads
 \begin{equation}
     3H(1+Q)\dot{\phi}+V_{,\phi}\simeq 0.
 \end{equation}
 Finally, we shall conclude this section by stating that
 some existing works \cite{kamali2016tachyonwarmintermediateinflationlight,Kamali_2020} have considered warm inflation in the RS-II brane which confronts the model with Planck data.
 \section{Inflationary Parameters In The RS-II Model} \label{Section 3}
We begin by further simplifying the Friedmann equation \eqref{f1} and neglect
the effective cosmological constant  $\Lambda_5$ and the extrinsic curvature $K$. Neglecting the extrinsic curvature is motivated by our observation of a very small $K$ in the universe. Moreover, its not necessarily true that the extrinsic curvature is small before and during standard inflation but, by the time a USR phase prevails, inflation would have gone through many e-folds, thus, suppressing the curvature by the start of the  USR phase. Moreover, we are justified in neglecting the effective cosmological constant $\Lambda _5$ since during inflation, the inflaton has a much bigger contribution in the accelerated expansion of the universe compared to the cosmological constant. 
\\
\\
Although the dark radiation term $\frac{\rho_{E_0}}{3M_{Pl}^2}\left(\frac{a_0^4}{a^4}\right)$ in eqn. \eqref{f1} rapidly redshifts away with the expansion of the universe, its inclusion is conceptually and physically important in a braneworld setting. This term arises from the projection of the bulk Weyl tensor onto the brane and encodes non-local gravitational effects from the higher-dimensional bulk geometry. Retaining it ensures that the full imprint of the five-dimensional dynamics on the brane is preserved, even if its numerical impact becomes subdominant during inflation. Moreover, the presence of dark radiation allows one to quantify precisely how bulk–brane energy exchange and initial conditions in the bulk might influence early-universe observables, such as the power spectrum or primordial black hole abundance. Finally, observational bounds from Big Bang Nucleosynthesis and the CMB still permit a small, nonzero dark radiation contribution, motivating its inclusion for completeness and consistency with the full RS-II framework (see for example \cite{Sasankan_2017}) .
With the above simplifications, the Friedmann equation takes the form
\begin{equation} \label{friedmann 1}
    H^2=\frac{1}{3M_{Pl}^2}\left[\frac{\dot{\phi}^2}{2}+V(\phi)+\frac{1}{2\lambda}\left(\frac{\dot{\phi}^4}{4}+V(\phi)\dot{\phi}^2+V^2(\phi)\right)+\rho_r+\frac{\rho_r}{2\lambda}\left(\rho_r+\dot{\phi}^2+2V(\phi)\right)\right]+\frac{\rho_{E_0}}{3M_{Pl}^2}\left(\frac{a_0}{a}\right)^4.
\end{equation}
Taking the derivative of the above expression with respect to time, we have
\begin{equation} \label{freidmann 2}
    2H\dot{H}=\frac{1}{3M_{Pl}^2}\left(\dot{\phi}\ddot{\phi}+V_{,\phi}+\frac{1}{2\lambda}(\dot{\phi}^3\ddot{\phi}+2\dot{\phi}\ddot{\phi}V(\phi)+V_{,\phi}(\phi)\dot{\phi}^2+2V(\phi)V_{,\phi}(\phi)+\frac{\rho_r}{\lambda}\dot{\phi}(\ddot{\phi}+V_{,\phi})\right)-\frac{4\rho_{E_0}a_0^4\dot{a}}{3M_{Pl}^2a^5}.
\end{equation}
Where we used the fact that $\dot{\rho}_r\simeq 0$. Now, by eqns. \eqref{Raddis} and \eqref{diss} we have 
\begin{equation}\label{rad}
    \rho_r=\frac{3}{4}Q\dot{\phi}^2
\end{equation}
In the inflationary approximation $V(\phi)>>\dot{\phi}^2$, from eqns. \eqref{friedmann 1} and \eqref{freidmann 2}, we arrive at the following equations
\begin{equation}
    H^2=\frac{V(\phi)}{3M_{Pl}^2}\left(1+\frac{V(\phi)}{2\lambda}\right)+\frac{\rho_{E_0}}{3M_{Pl}^2}\left(\frac{a_0}{a}\right)^4
\end{equation}
and
\begin{equation}
    \dot{H}=\frac{1}{6HM_{Pl}^2}\left(\dot{\phi}\ddot{\phi}+\frac{1}{2\lambda}(\dot{\phi}^3\ddot{\phi}+2\dot{\phi}\ddot{\phi}V(\phi)\right)-\frac{2\rho_{E_0}a_0^4}{3M_{Pl}^2a^4}\simeq\frac{\dot{\phi}\ddot{\phi}}{6HM_{Pl}^2}\left(1+\frac{V(\phi)}{\lambda}\right)-\frac{2\rho_{E_0}a_0^4}{3M_{Pl}^2a^4}.
\end{equation}
By the equation of motion $\ddot{\phi}+3H\dot{\phi}(1+Q)\simeq 0$, we get 
\begin{equation}
    \dot{H}=\frac{-(1+Q)\dot{\phi}^2}{2M_{Pl}^2}\left[1+\frac{1}{\lambda}\left(\frac{\dot{\phi}^2}{2}+V(\phi)\right)\right]-\frac{2\rho_{E_0}a_0^4}{3M_{Pl}^2a^4}\simeq\frac{-(1+Q)\dot{\phi}^2}{2M_{Pl}^2}\left[1+\frac{V(\phi)}{\lambda}\right]-\frac{2\rho_{E_0}a_0^4}{3M_{Pl}^2a^4} .
\end{equation}
In order to simplify our expressions, we shall introduce parameters $\alpha$ and $\beta$ as
\begin{equation} \label{alpha}
    \alpha=\frac{V(\phi)}{2\lambda}
\end{equation}
\begin{equation} \label{beta}
 \beta=\frac{\rho_{E_0}a_0^4}{3a^4}.
\end{equation}
Here $\alpha$ is the dimensionless ratio between the potential and the brane tension and plays the role of resetting the strength of the potentials relative to the brane tension. Furthermore, $\beta$ signifies the role of dark radiation in modifying the Friedmann equations. Observational limits from Big Bang Nucleosynthesis and then $CMB$ constrain the dark radiation term $\beta$ in RS-II models to be of order a few percent of the energy density, with recent bounds indicating $\frac{\rho_{DR}}{\rho}\in [-6\%,+6\%]$ \cite{Sasankan_2017}. For example, at high energies; $V>>\lambda$ and hence, $\alpha>>1$ where as, for low energies, $\alpha<<1$. The value $\alpha=0$ corresponds to a general relativistic background cosmology. The parameter $\beta$ quantifies the role of dark radiation from the bulk to the Friedmann equations. 
We subsequently proceed to calculate inflationary parameters like $\epsilon_1$ and $\epsilon_2$. To start off , from the expressions of $\dot{H}$ and $\ddot{H}$, we can derive that $\epsilon_1$ takes the following form
\begin{equation}
    \epsilon_1=\frac{3}{2}\left[\frac{(1+Q)\dot{\phi}^2(1+2\alpha)+2\beta}{V(\phi)(1+\alpha)+\beta}\right].
\end{equation}
Now, 
\begin{equation}
    \epsilon_2=\frac{\ddot{H}}{H\dot{H}}+2\epsilon_1
\end{equation}
and hence, in order to calculate $\epsilon_2$, we first proceed with calculating $\frac{\ddot{H}}{H\dot{H}}$ and hence, first calculate $\ddot{H}.$
\begin{align}
    \ddot{H}=-\frac{\dot{Q}\dot{\phi}^2+2Q\dot{\phi}\ddot{\phi}}{2M^2_{PI}}\left[1+\frac{1}{\lambda}\left(\frac{\dot{\phi}^2}{2}+V(\phi)\right)\right]-\frac{(1+Q)\dot{\phi}^3\ddot{\phi}}{2M^2_{PI}\lambda}+\frac{8\rho_{E_0}a_0^4H}{3M^2_{PI}a^4}\\
 =\frac{-\dot{\phi}^2(\dot{Q}-6HQ(1+Q))(1+2\alpha)}{2M_{Pl}^2}+\frac{8\beta H}{2M^2_{PI}}+\frac{3H(1+Q)^2\dot{\phi}^4}{2M_{Pl}^2\lambda}.
\end{align}
Finally, we have 
\begin{equation}
    \frac{\ddot{H}}{H\dot{H}}=-\frac{3}{2}\left[\frac{\dot{\phi}^2(6Q(1+Q))(1+2\alpha)+8\beta+3(1+Q)^2\frac{\dot{\phi}^4}{\lambda}}{(1+  Q)\dot{\phi}^2(1+2\alpha)+2\beta}\right]-\frac{6M_{Pl}\dot{Q}\dot{\phi}^2\sqrt{3}}{[(1+  Q)\dot{\phi}^2(1+2\alpha)+2\beta]\sqrt{V(\phi)(1+\alpha)+\beta}}.
\end{equation}
and subsequently, we arrive at the following important result
\begin{align}   \epsilon_2=3\left[\frac{(1+Q)\dot{\phi}^2(1+2\alpha)+2\beta}{V(\phi)(1+\alpha)+\beta}\right]-\frac{3}{2}\left[\frac{\dot{\phi}^2(6Q(1+Q))(1+2\alpha)+8\beta+3(1+Q)^2\frac{\dot{\phi}^4}{\lambda}}{(1+  Q)\dot{\phi}^2(1+2\alpha)+2\beta}\right]\notag \\-\frac{6M_{Pl}\dot{Q}\dot{\phi}^2\sqrt{3}}{[(1+  Q)\dot{\phi}^2(1+2\alpha)+2\beta]\sqrt{V(\phi)(1+\alpha)+\beta}}.
\end{align}
It is convenient to further introduce parameters $\gamma$ and $\delta $ as
\begin{equation}\label{gamma}
    \gamma=\frac{V}{\dot{\phi}^2}.
\end{equation}
and
\begin{equation}
    \delta=\frac{\beta}{\dot{\phi}^2}.
\end{equation}
For clarity, we shall now summarize various parameters introduced in the text .  The parameter $\alpha$, given by eqn.\eqref{alpha}, physically represents high-energy corrections induced due to the  brane tension. The parameter $\beta$, given by eqn.\eqref{beta}, signifies correction induced due to dark-radiation from the bulk and is more precisely, equal to $\rho_{E_0}a_0^4/3a^4$. The dimensionless  parameter  $Q=\Gamma/3H$ measures the ratio of two friction terms appearing in the equation of motion for $\phi$, one due to the dissipation and the other the other due to the expansion of the universe. The parameter $\gamma$ satisfies the following relation $2\gamma=\frac{V}{\left(\frac{\dot{\phi}^2}{2}\right)}$ which is the ratio  between the potential and kinetic energy of the inflaton $\phi$. The parameter $\delta$ evaluates to $\beta/\dot{\phi}^2$ which is the ratio between dark radiation and  the kinetic energy of the inflaton  $\phi$.
\\
\\
We now proceed to obtain certain concrete and observational quantities such as the number of e-folds and the power spectrum for curvature perturbations during the USR phase of warm inflation inflation in a braneworld background cosmology. 
\\
\\
The number of e-folds prevailed as the inflaon evolves takes the following form
\begin{equation}
    N_e=\int d\phi \frac{H(\phi)}{\dot{\phi}} =\frac{1}{M_{Pl}\sqrt{3}}\int  d\phi \sqrt{\gamma(1+\alpha)+\delta}.
\end{equation}
Now, from texts like \cite{Bastero_Gil_2019,Hall2004,Moss2008,BasteroGil2011,Graham2009}; we know that for warm inflation, the power spectrum gains additional corrections as follows
\begin{equation}\label{P2}
    P^2=\left(\frac{H^2}{2\pi\dot{\phi}^2}\right)^2 [1+C(Q,T,..)]
\end{equation}
where $C(Q,T...)$ is an arbitrary function whose form depends on the nature of the dissipative mechanism. The factor $1+C(Q,T,..)$ does not depend on braneworld corrections and hence. Our purpose here  is to analyze the effect of braneworld corrections to inflationary cosmology. Hence, while performing a numerical analysis in section \ref{Section 4}, we shall only content with ourselves with analyzing the first term $\left(\frac{H^2}{2\pi\dot{\phi}^2}\right)^2$ and then judge the effects of braneworld corrections to the primordial power spectrum. Note that the closed form expression for $\left(\frac{H^2}{2\pi\dot{\phi}^2}\right)^2$ is given by
\begin{equation}\label{P2}
    \left(\frac{H^2}{2\pi\dot{\phi}^2}\right)^2=\left(\frac{1}{6\pi M_{Pl}}\right)^2\left(\gamma(1+\alpha)+\delta\right)^2.
\end{equation}

With this, we shall conclude this section where we have investigated the USR phase of warm inflation in a background RS-II cosmology. In the next section, we analyse the the USR phase of warm inflation in the DGP setup. 
\section{Inflationary Parameters In The DGP Model}\label{Section 4}

Recall from section \ref{Section 2} that the effective Friedmann equation for DGP cosmology takes the following form
\begin{equation} \label{DGP1}
    H=\sqrt{\frac{\rho}{3M_{Pl}^2}+\frac{1}{4r_c^2}}+\frac{\epsilon}{2r_c}
\end{equation}
where $\rho=\rho_{\phi}+\rho_r=\frac{\dot{\phi}^2}{2}(1+1.5Q)+V(\phi)$ and $\epsilon=\pm 1$ labels two possible branches: the self-accelerating branch ($\epsilon=+1$) and the normal branch ($\epsilon=-1$). In our analysis, we will not choose a particular branch but instead have a generic $\epsilon=\pm 1$. One can always choose a particular branch by making a suitable choice for $\epsilon$. 
\\
\\
Differentiating eqn. \eqref{DGP1} with respect to time now yields,
\begin{equation} \label{DGP2}
    \dot{H}=\frac{\dot{\rho}}{6M_{Pl}^2\sqrt{\frac{\rho}{3M_{Pl}^2}+\frac{1}{4r_c^2}}}=\frac{-H(1+Q)\dot{\phi}^2}{2M_{Pl}^2\sqrt{\frac{\rho}{3M_{Pl}^2}+\frac{1}{4r_c^2}}}
\end{equation}
The first slow roll parameter now takes the form 
\begin{equation} \label{sl 1}
    \epsilon_1=\frac{-\dot{H}}{H^2}=\frac{(1+Q)\dot{\phi}^2}{2M_{Pl}^2H \sqrt{\frac{\rho}{3M_{Pl}^2}+\frac{1}{4r_c^2}}}.
\end{equation}
\\
\\
Now, in the warm inflationary model, the background radiation density remains constant:-
\begin{equation} \label{Raditaion constanticy}
    \dot{\rho_r}\simeq0.
\end{equation}
This means that $\dot{\rho}=\dot{\rho}_r+\dot{\rho}_{\phi}=\dot{\rho}_{\phi}=-3H\dot{\phi}^2(1+Q).$ Eqns. \eqref{Raditaion constanticy} and \eqref{rad} together imply that
\begin{equation} \label{Q}
    \frac{d(Q\dot{\phi}^2)}{dt}=0.
\end{equation}
We remark that the above eqn. \eqref{Q} can also be written as
\begin{equation}
    \dot{Q}=-6  HQ(1+Q)
\end{equation}
this form will be particularly useful in making plots and performing the numerical analysis in \ref{Section 4}.
\\\
Returning back to the $DGP$ model,we differentiate eqn. \eqref{DGP2} and use eqn. \eqref{Q} to get,
\begin{equation}
    \ddot{H}=\frac{-\dot{H}(1+Q)\dot{\phi}^2+6H^2(1+Q)^2\dot{\phi}^2}{2M_{Pl}^2\sqrt{\frac{\rho}{3M_{Pl}^2}+\frac{1}{4r_c^2}}}-\frac{H^2(1+Q)^2\dot{\phi}^4}{4M_{Pl}^4\left(\frac{\rho}{3M_{Pl}^2}+\frac{1}{4r_c^2}\right)^{3/2}}=\frac{\dot{H}^2}{H}-6\dot{H}H-\frac{\dot{H}^2}{H-\frac{\epsilon}{2r_c}}.
\end{equation}
The second slow roll parameter is now given by
\begin{equation} \label{sl2}
    \epsilon_2=\frac{\ddot{H}}{H\dot{H}}+2\epsilon_1=-\epsilon_1-6+\frac{\epsilon_1}{1-\frac{\epsilon}{2r_cH}}+2\epsilon_1=-\epsilon_1\frac{4r_cH- \epsilon}{2r_cH-\epsilon}-6.
\end{equation}
As in section \ref{Section 3}, we proceed to obtain concrete and observational quantities such as the number of e-folds and the power spectrum for curvature perturbations during the USR phase of warm inflation. In the inflationary approximation ($V(\phi)>>\dot{\phi}^2$), the number of e-folds prevailed as the inflaton evolves are given by
\begin{equation}
    N_e=\int d\phi \frac{H}{\dot{\phi}}=\frac{1}{M_{Pl}\sqrt{3}}\int d\phi \left[\sqrt{\gamma+\chi^2}+\epsilon\chi\right]
\end{equation}
where $\gamma$ is given as in eqn. \eqref{gamma} and $\chi$ is given as 
\begin{equation}
    \chi=\frac{M_{Pl}\sqrt{3}}{2r_c\dot{\phi}}.
\end{equation}
Subsequently, the power spectrum for curvature perturbations takes the following form
\begin{equation}
P^2=\left(\frac{\left(\sqrt{\gamma+\chi^2}+\epsilon\chi\right)^2}{6\pi M_{Pl}^2}\right)^2[1+C(Q,T,..)].
\end{equation}
where $C(Q,T,..)$ is an arbitrary function which arises due to dissipation  and depends on the dissipative mechanism as described in section \ref{Section 2}.
\\
\\
After performing the above analysis and finding a closed form expression for the slow roll parameters, the number of e-folds the system undergoes and the power spectrum of scalar curvature perturbations, we conclude that the DGP model would not have significant effects during the USR phase of warm inflation too. This is because during inflation, $\frac{\rho}{3M_{Pl}^2}>>r_c$ and $\gamma>>\chi$. As seen by eqns. \eqref{sl 1} and \eqref{sl2}, the slow roll parameters all reduce to their familiar expressions in an $FLRW$ based USR phase of warm inflation in the limit $r_c\rightarrow \infty$. Furthermore, the DGP induced $\chi$ contributions to the primordial power spectrum of scalar curvature perturbations don't have a significant role  since $\gamma>>\chi$ during inflation. This goes in hand with the general well known statement that the DGP model induces lower effects during early universe cosmology as compared to the RS-II braneworld model which generally induces non-negligible effects during of early universe cosmology. In this section we concluded that DGP model doesn't have significant effects even when the inflaton enters a USR phase. It remains to be seen whether RS-II braneworld effects still remain subdominant during a USR phase which we shall do in the next section. 

\section{Numerical Results} \label{Section 5}
\begin{figure*}
    \centering
    \begin{subfigure}[b]{0.48\textwidth}
        \centering
        \includegraphics[width=\linewidth]{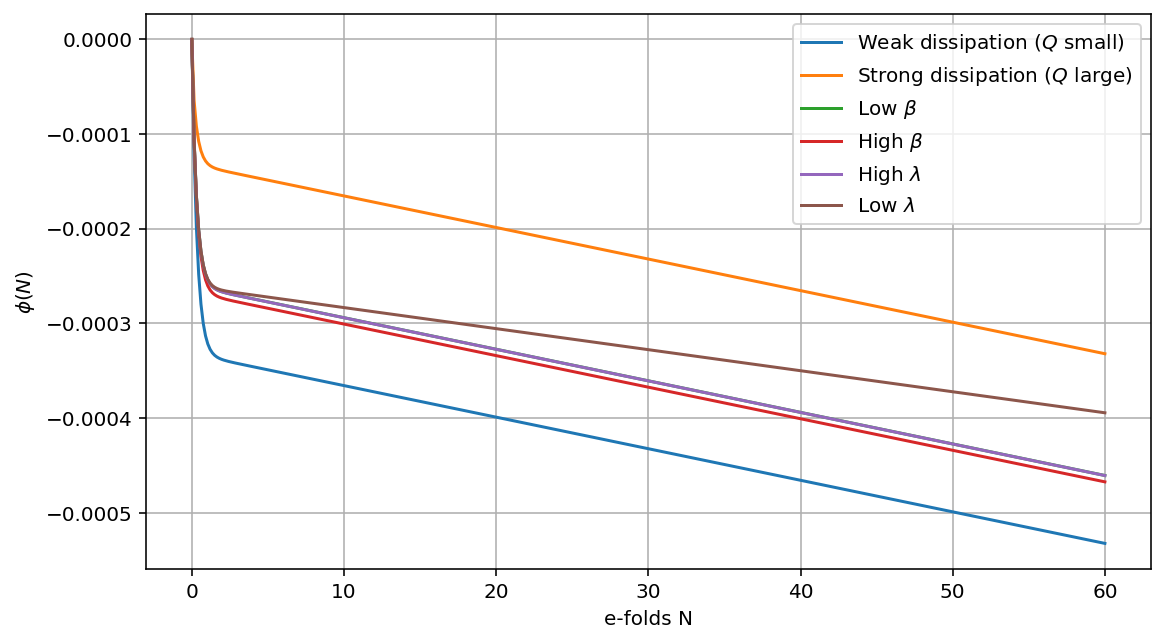}
        \caption{Evolution of $\phi$ with e-folds for a linearized potential at $\frac{M}{V_0}=10^{-5}$. Base values: $Q_0=1$, $\rho_{E0}=3M^2_{Pl}10^{-3}$, $a_0=1$, $\lambda=10^{10}$. Weak dissipation: $Q_0=10^{-6}$; strong dissipation: $Q_0=10$. High $\beta$: $\rho_{E_0}=3M_{Pl}10^{-1}$; low $\beta$: $\rho_{E_0}=3M_{Pl}10^{-12}$. High $\lambda$: $\lambda=10^{50}M^2_{Pl}$; low $\lambda$: $\lambda=M_{Pl}^2$.}
        \label{fig:sub1}
    \end{subfigure}
    \hfill
    \begin{subfigure}[b]{0.48\textwidth}
        \centering
        \includegraphics[width=\linewidth]{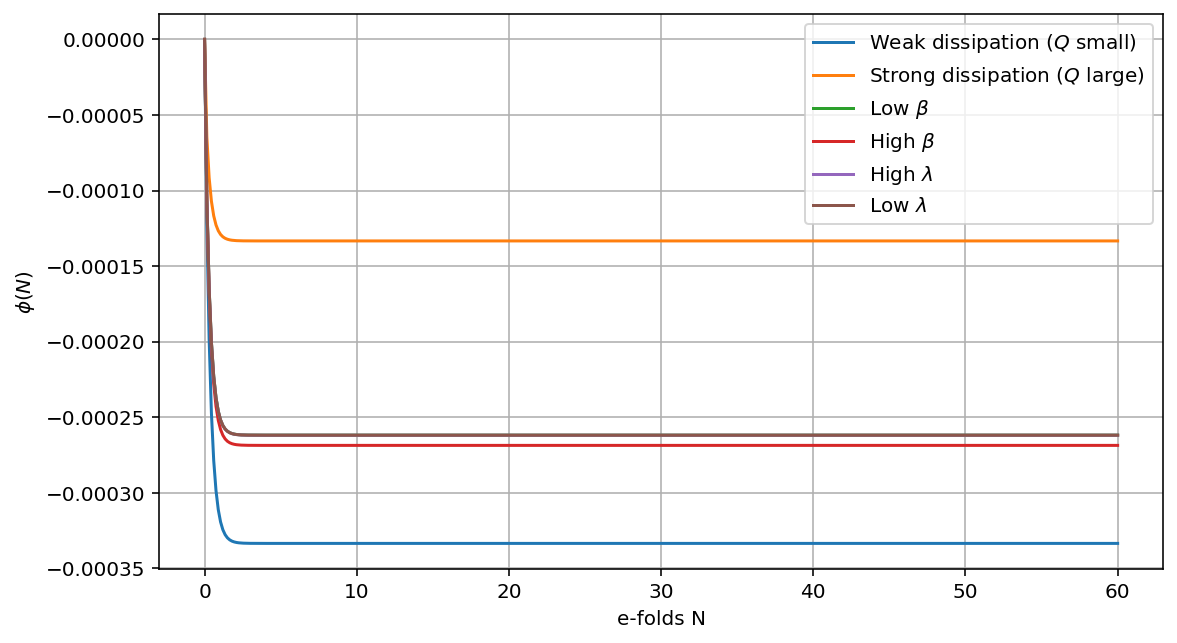}
        \caption{Inflaton $\phi$ for the linearized potential with same parameters as (a) but for $M\rightarrow 0$.}
        \label{fig:sub2}
    \end{subfigure}

    \begin{subfigure}[b]{0.48\textwidth}
        \centering
        \includegraphics[width=\linewidth]{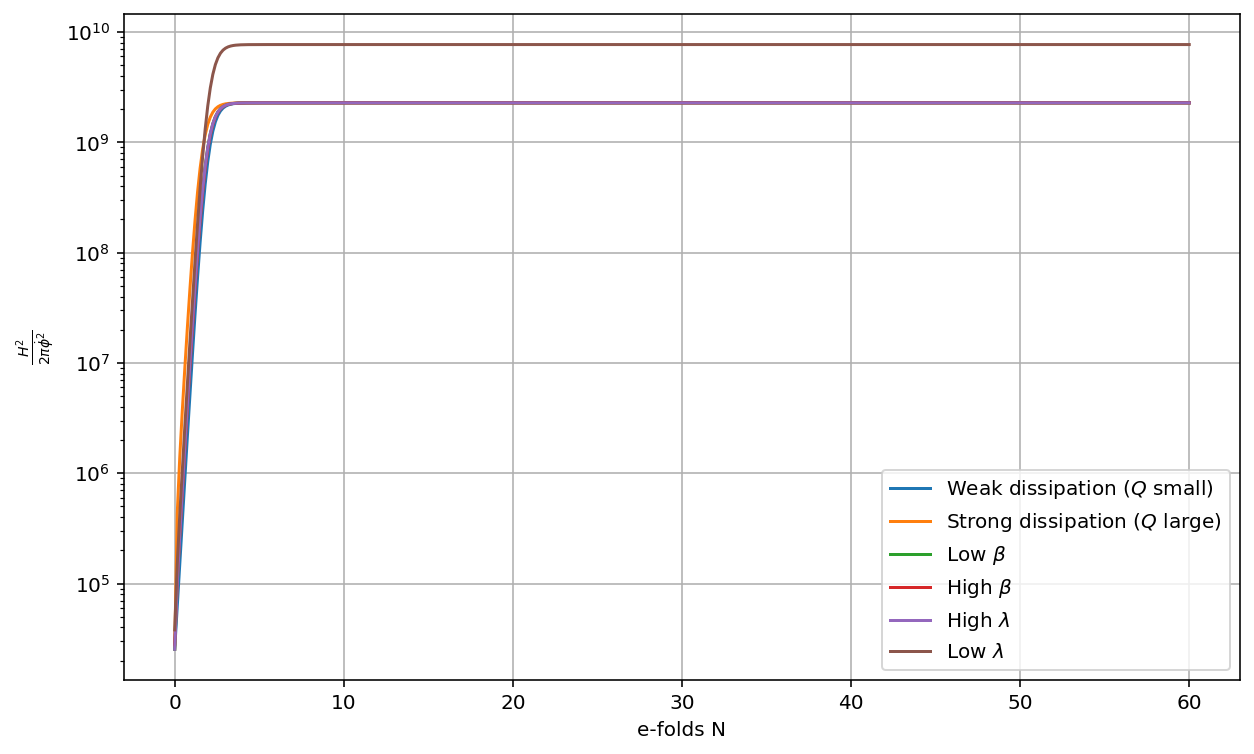}
        \caption{Primordial power spectrum for $\frac{M}{V(\phi)}=10^{-5}$ in a linearized potential with same parameters as Figure~\ref{fig:sub1}.}
        \label{fig:sub3}
    \end{subfigure}
    \hfill
    \begin{subfigure}[b]{0.48\textwidth}
        \centering
        \includegraphics[width=\linewidth]{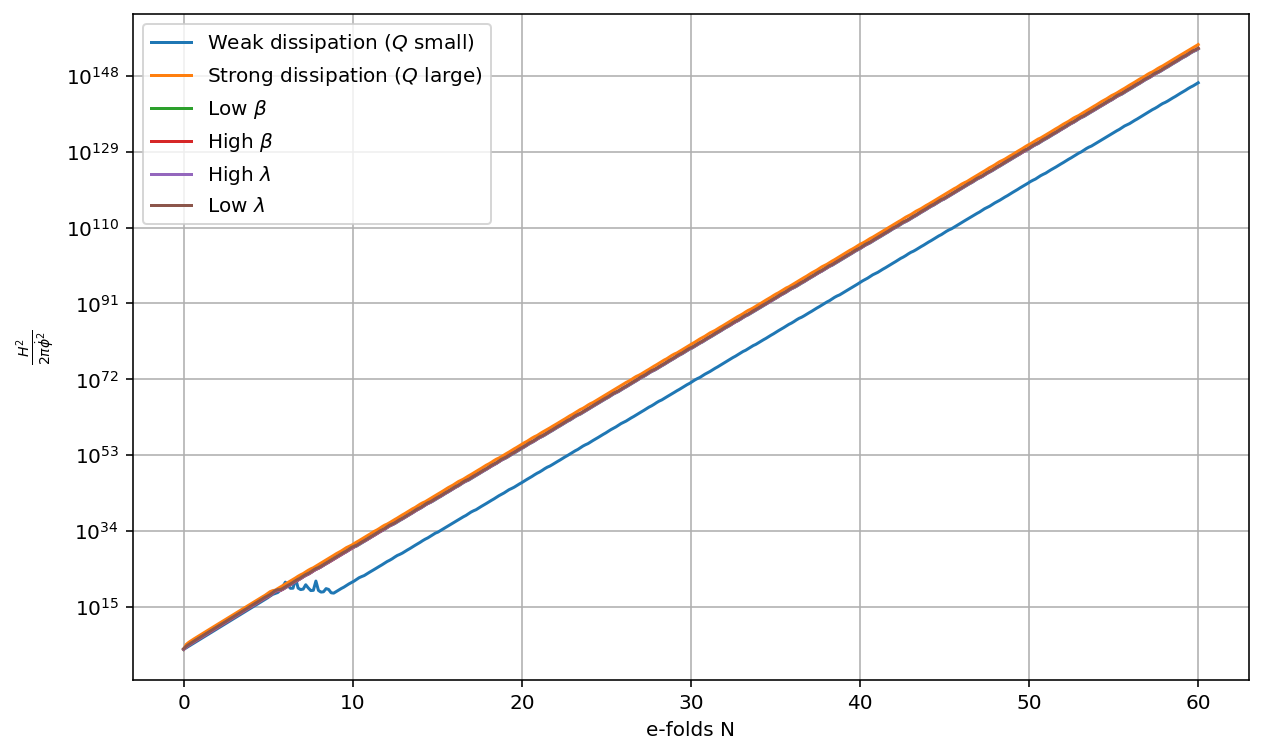}
        \caption{Primordial power spectrum for $M\rightarrow 0$ in a linearized potential.}
        \label{fig:sub4}
    \end{subfigure}

    \begin{subfigure}[b]{0.48\textwidth}
        \centering
        \includegraphics[width=\linewidth]{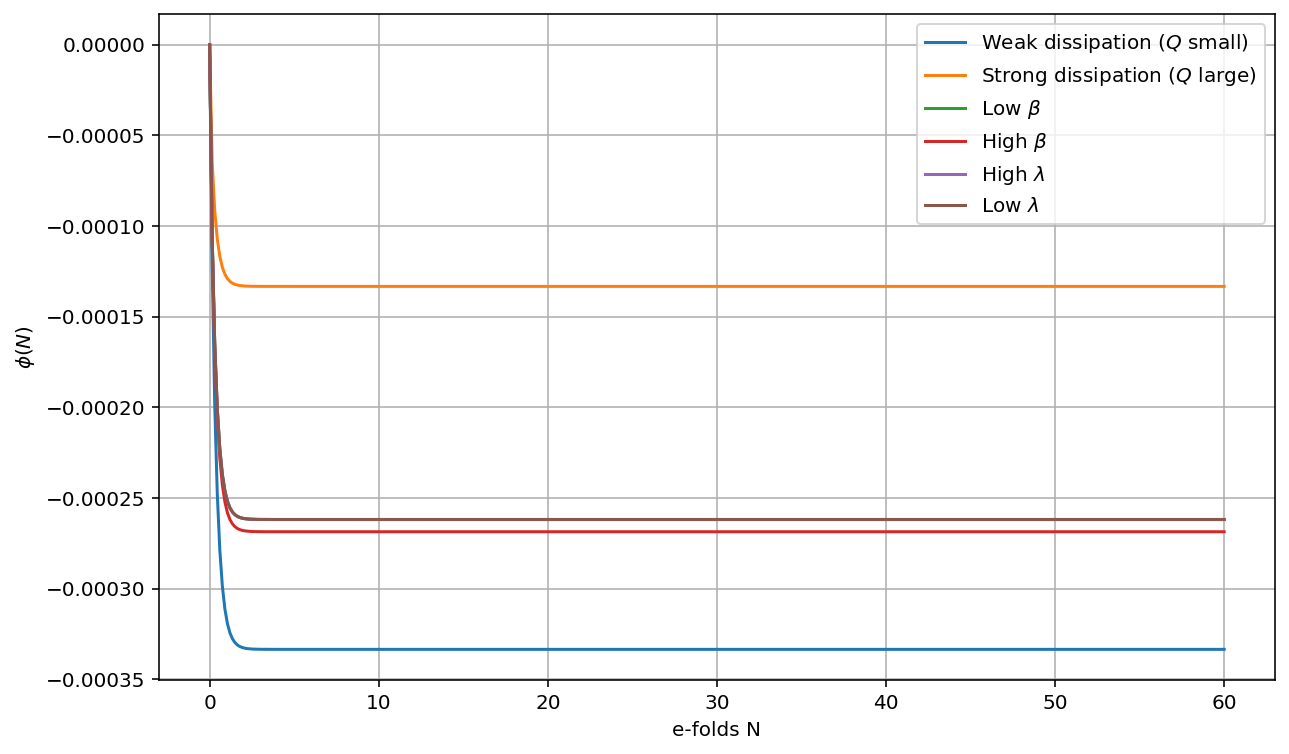}
        \caption{Inflaton $\phi$ in a cubic potential with $\frac{M}{V_0}=10^{-5}$.}
        \label{fig:sub5}
    \end{subfigure}
    \hfill
    \begin{subfigure}[b]{0.48\textwidth}
        \centering
        \includegraphics[width=\linewidth]{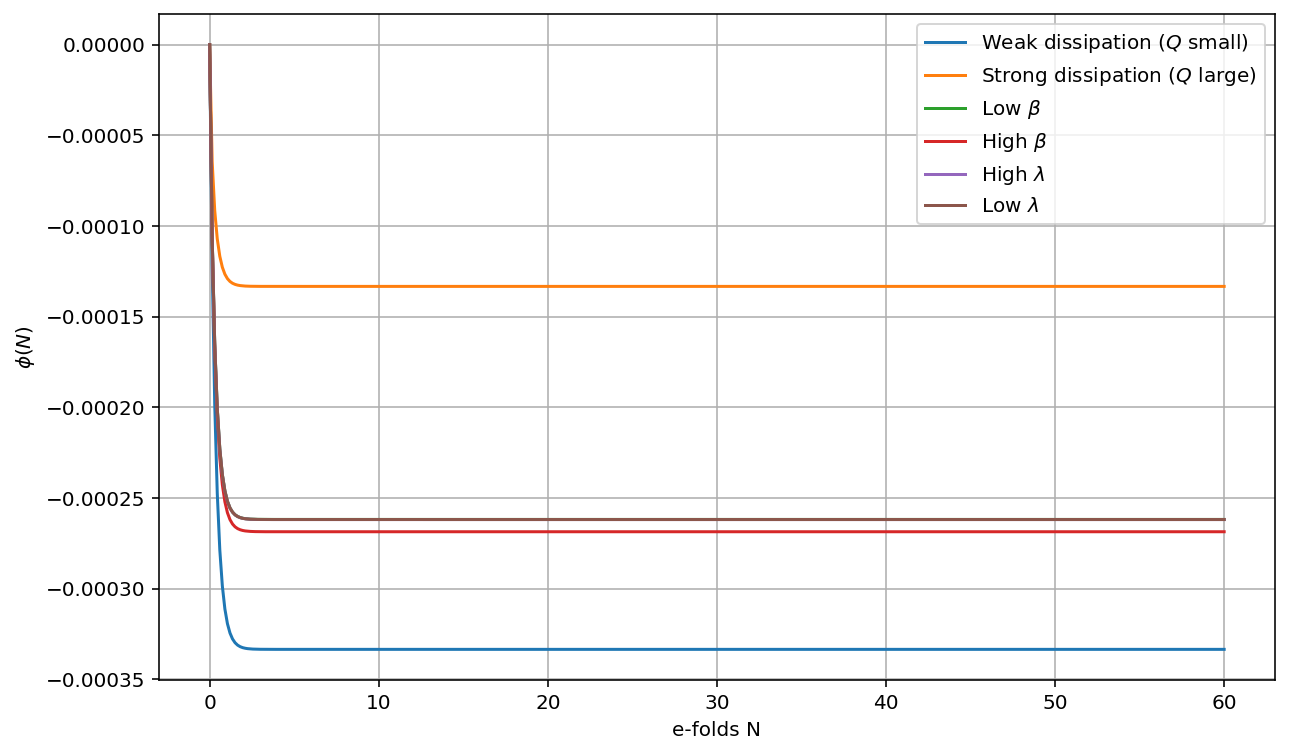}
        \caption{Inflaton $\phi$ in a cubic potential with $M\rightarrow 0$.}
        \label{fig:sub6}
    \end{subfigure}

    \begin{subfigure}[b]{0.48\textwidth}
        \centering
        \includegraphics[width=\linewidth]{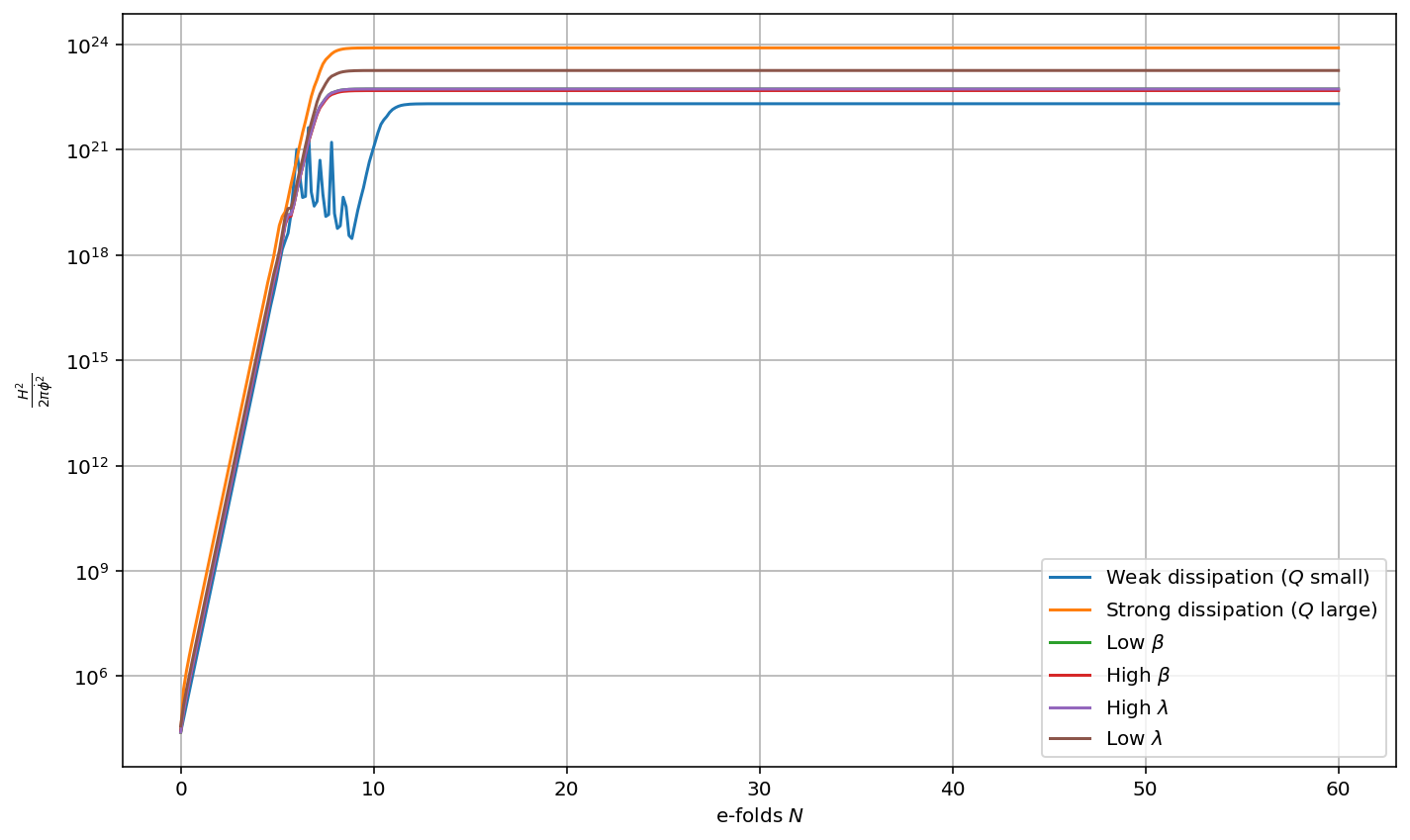}
        \caption{Primordial power spectrum $P$ in a cubic potential with $\frac{M}{V_0}=10^{-5}$.}
        \label{fig:sub7}
    \end{subfigure}
    \hfill
    \begin{subfigure}[b]{0.48\textwidth}
        \centering
        \includegraphics[width=\linewidth]{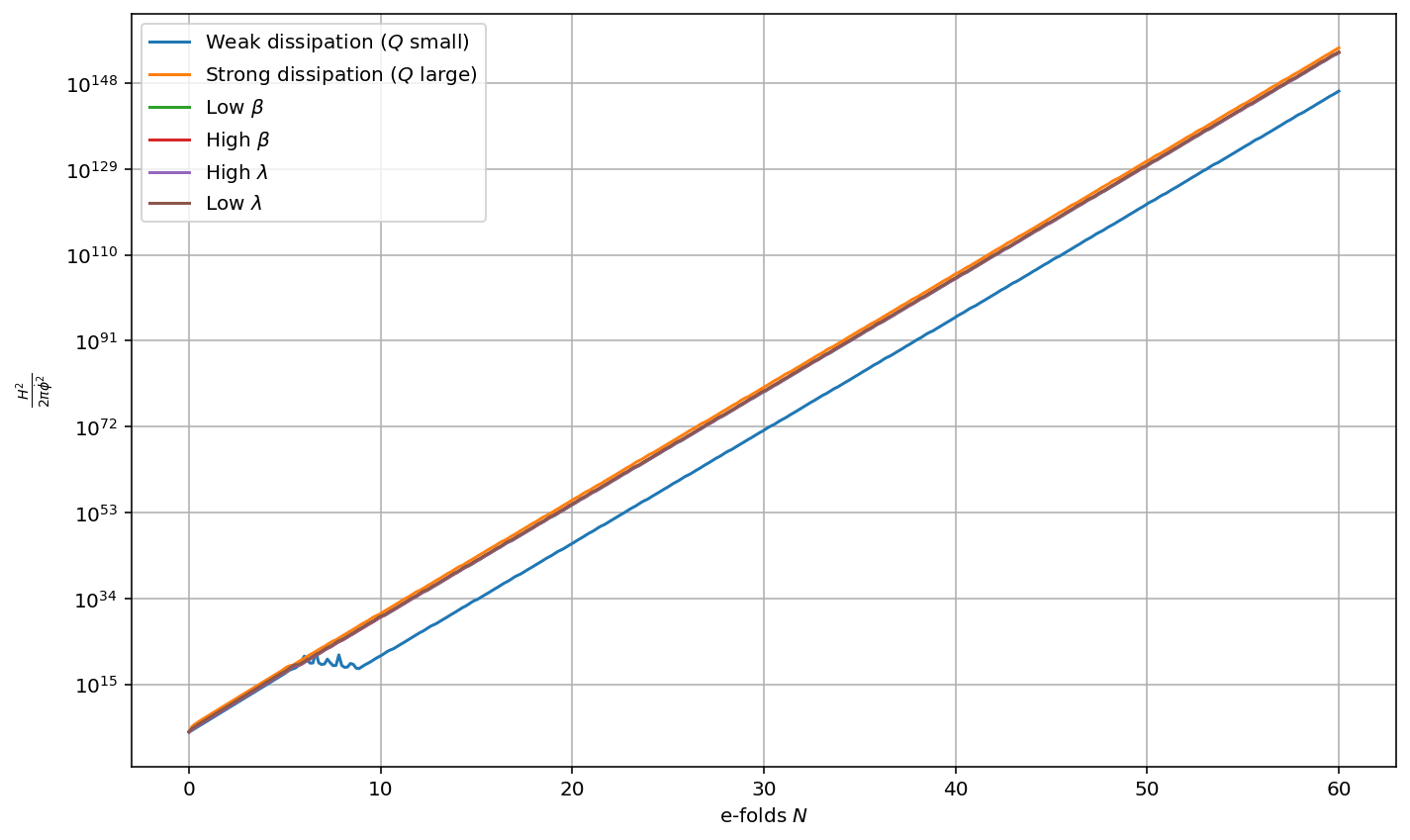}
        \caption{Primordial power spectrum $P$ in a cubic potential with $M\rightarrow 0$.}
        \label{fig:sub8}
    \end{subfigure}

    \caption{Comparison of inflaton evolution and primordial power spectra for linearized and cubic potentials under different values of $M$. Each subplot illustrates the dependence of the inflaton field $\phi$ and the primordial power spectrum $P$ on dissipation, brane tension, and dark radiation parameters.}
    \label{fig:combined}
\end{figure*}

We will now numerically solve for the evolution of the inflaton $\phi$ and the primordial power spectrum for scalar curvature perturbations $P$ in cases where warm inflation can enter an USR phase in a background $RS-II$ braneworld cosmology, and then appreciate its characteristics. We will consider two different potentials, the linear potential in and the cubic potential .
\\
\begin{itemize}
\item \textbf{Linear Potential:-}
To start , we shall consider the case of a linearized potential $V(\phi)=V_0+M_0\phi$.
In this potential, when $V_0>>M_0\phi$, the potential becomes extremely flat, and the system can enter a USR phase. Figures \ref{fig:sub1} and  \ref{fig:sub2}, show the background evolution of $\phi(N) $ for the linear potential.
The dominant effect is the initial value of the dissipation parameter $Q_0$,  increasing $Q_0$ from $10^{-6}$ to $10$ delays the field’s descent and reduces the inflaton $\phi$ significantly. One can immediately see by comparing figures \ref{fig:sub1} and \ref{fig:sub2} that the effects braneworld induced corrections reduce significantly when one is stricter about the USR condition and eventually become negligible when one is very strict about USR.
Furthermore, by comparing fig.(\ref{fig:sub1}) with (\ref{fig:sub2}), we see that the more stricter we are about USR, the flatter the inflaton $\phi$ gets during late times (a higher number of e-folds have passed).
\\
\\
Figs.(\ref{fig:sub3}) and (\ref{fig:sub4}) show that being strict about the USR phase significantly changes the form of the primordial power spectrum for scalar curvature perturbations (recall that the form of $\left(\frac{H^2}{2\pi\dot{\phi}^2}\right)^2$ directly influences the form of the primordial power spectrum even when you include dissipative effects as in warm inflation). The graphs given in fig. (\ref{fig:sub4}) reflect that braneworld induced terms have a low effect while the initial value of the dissipation coefficient has a significant effect on the power spectrum while when one is more relaxed about USR (as in fig. (\ref{fig:sub3})),one sees that the brane tension have significant effect on the power spectrum as compared to the initial value of the dissipative coefficient. 
\\
\\
To conclude, the evolution of $\phi$ remains invariant by increasing the brane tension $\lambda$. However, one sees that the brane tension $\lambda$ does have a sub-dominant effect on the curvature perturbations. 
\item \textbf{Cubic Potential:-}
We shall now deal with the potential $V=V_0+M_{Pl}\phi^3$.  In fig. (\ref{fig:sub6}) and fig. (\ref{fig:sub7}), we see that the effect on the inflation  $\phi$ of changing from $M=10^{-5}V_0$ to $M\rightarrow 0$ is little smaller compared to that of a linearized potential. The effect of varying the brane tension $\lambda$ is even more suppressed in the cubic potential case, while the dark radiation coming from the bulk still has a small subdominant effect on the inflaton $\phi$ even in the case of a cubic potential.
\\
\\
Furthermore, comparing  fig. (\ref{fig:sub8}) with fig. (\ref{fig:sub3}) and fig. (\ref{fig:sub4}) shows  that the evolution of the power spectrum is similar for cubic and linearized potentials in the $M\rightarrow 0$ limit. However, a cubic potential shows visibly distinct behavior than a linear potential for the case $M=10^{-5}$.

\end{itemize}
\newpage
\section{Conclusions}\label{Conclusion}

In this work, we have studied the embedding of an USR phase within the warm inflation framework in the context of the Randall-Sundrum type II (RS-II) braneworld cosmology. Starting from the modified Friedmann equations, we derived the background dynamics, generalized slow-roll parameters, and the curvature power spectrum, explicitly incorporating the dissipation ratio $Q$, the brane tension parameter $\alpha$ and the dark radiation contribution $\beta$.
\\
\\
Our analytical and numerical analyses reveal that while the RS-II braneworld framework introduces formal high-energy corrections to the Friedmann equations, its phenomenological effects are virtually negligible once the system enters the USR phase of warm inflation. This indicates that USR dynamics in warm inflation operates almost entirely within the standard four-dimensional dynamics, with RS-II braneworld induced terms contributing only as tiny higher-order corrections.
\\
\\
Moreover, we then studied the $USR$ phase of warm inflation in a background DGP model. The DGP model in general doesn't have significant effects during early universe cosmology and our study showed that this fact didn't change even when the inflaton entered a USR phase. It seems that braneworld effects in general are \textbf{not} very significant during the USR phase of warm inflation.  
\\
\\
Future work could extend this analysis by considering other high-energy scenarios (such as Gauss-Bonnet models) to test whether this sub-dominance is a universal feature of ultra slow roll dynamics of warm inflation. Furthermore, incorporating Non-Gaussian effects \cite{Byrnes2012_PBH_NG_Constraint,YoungMuscoByrnes2019_NonlinearRelation,young2024computingabundanceprimordialblack,KehagiasMuscoRiotto2019_Threshold,DeLuca2019_IneludibleNG,Gow2022_NonPerturbative,pi2025primordialblackholeformation,Pi2024_ReviewNG_PBH,Pi_2025,Montefalcone2023} and full PBH mass functions would refine predictions for PBH abundances \cite{young2024computingabundanceprimordialblack,Motohashi_2017,pi2025primordialblackholeformation,Arya2019} within this framework.
\section{Acknowledgments}

I would like to express my deepest gratitude to Oem Trivedi for his invaluable guidance, mentorship, and encouragement throughout the course of this research. His expertise in the areas of warm inflationary cosmology and USR dynamics was instrumental in shaping both the conceptual foundation and the development of this work.
\\
\\
I am also sincerely thankful to Meet Vyas, Kanabar Jay and Maxim Khlopov for their insightful discussions and constructive feedback, which helped refine the presentation and broaden the context of this manuscript.
\bibliographystyle{unsrt}
\bibliography{references}

\begin{thebibliography}{10}

\bibitem{Baumann:2022cosmology}
Daniel Baumann.
\newblock {\em Cosmology}.
\newblock Cambridge University Press, 2022.

\bibitem{Mukhanov:2005fem}
Viatcheslav Mukhanov.
\newblock {\em Physical Foundations of Cosmology}.
\newblock Cambridge University Press, 2005.

\bibitem{Weinberg:2008cosmology}
Steven Weinberg.
\newblock {\em Cosmology}.
\newblock Oxford University Press, 2008.

\bibitem{Dodelson:2003modern}
Scott Dodelson.
\newblock {\em Modern Cosmology}.
\newblock Academic Press, 2003.

\bibitem{Liddle:2000introduction}
Andrew~R. Liddle and David~H. Lyth.
\newblock {\em Cosmological Inflation and Large-Scale Structure}.
\newblock Cambridge University Press, 2000.

\bibitem{Kolb:1990early}
Edward~W. Kolb and Michael~S. Turner.
\newblock {\em The Early Universe}.
\newblock Addison-Wesley, 1990.

\bibitem{guth1981}
A.~H. Guth.
\newblock The inflationary universe: A possible solution to the horizon and flatness problems.
\newblock {\em Phys. Rev. D}, 23:347--356, 1981.

\bibitem{Linde:1982new}
Andrei~D. Linde.
\newblock A new inflationary universe scenario: A possible solution of the horizon, flatness, homogeneity, isotropy and primordial monopole problems.
\newblock {\em Phys. Lett. B}, 108:389--393, 1982.

\bibitem{Albrecht:1982eternal}
Andreas Albrecht and Paul~J. Steinhardt.
\newblock Cosmology for grand unified theories with radiatively induced symmetry breaking.
\newblock {\em Phys. Rev. Lett.}, 48:1220--1223, 1982.

\bibitem{Lyth:2009inflation}
David~H. Lyth and Andrew~R. Liddle.
\newblock {\em The Primordial Density Perturbation: Cosmology, Inflation and the Origin of Structure}.
\newblock Cambridge University Press, 2009.

\bibitem{Villanueva_Domingo_2021}
Pablo Villanueva-Domingo, Olga Mena, and Sergio Palomares-Ruiz.
\newblock A brief review on primordial black holes as dark matter.
\newblock {\em Frontiers in Astronomy and Space Sciences}, 8, May 2021.

\bibitem{kazanas1980}
D.~Kazanas.
\newblock Dynamics of the universe and spontaneous symmetry breaking.
\newblock {\em Astrophys. J. Lett.}, 241:L59--L63, 1980.

\bibitem{albrecht1982}
A.~Albrecht and P.~J. Steinhardt.
\newblock Cosmology for grand unified theories with radiatively induced symmetry breaking.
\newblock {\em Phys. Rev. Lett.}, 48:1220--1223, 1982.

\bibitem{linde1982}
A.~D. Linde.
\newblock A new inflationary universe scenario: A possible solution of the horizon, flatness, homogeneity, isotropy and primordial monopole problems.
\newblock {\em Phys. Lett. B}, 108:389--393, 1982.

\bibitem{sato1981b}
K.~Sato.
\newblock First order phase transition of a vacuum and expansion of the universe.
\newblock {\em Mon. Not. R. Astron. Soc.}, 195:467--479, 1981.
\newblock NORDITA-80-29.

\bibitem{sato1981a}
K.~Sato.
\newblock Cosmological baryon number domain structure and the first order phase transition of a vacuum.
\newblock {\em Phys. Lett. B}, 99:66--70, 1981.

\bibitem{young2024computingabundanceprimordialblack}
Sam Young.
\newblock Computing the abundance of primordial black holes, 2024.

\bibitem{pi2025primordialblackholeformation}
Shi Pi, Misao Sasaki, Volodymyr Takhistov, and Jianing Wang.
\newblock Primordial black hole formation from power spectrum with finite-width, 2025.

\bibitem{Motohashi_2017}
Hayato Motohashi and Wayne Hu.
\newblock Primordial black holes and slow-roll violation.
\newblock {\em Physical Review D}, 96(6), September 2017.

\bibitem{Carr_2016}
Bernard Carr, Florian Kühnel, and Marit Sandstad.
\newblock Primordial black holes as dark matter.
\newblock {\em Physical Review D}, 94(8), October 2016.

\bibitem{Berera_1995}
Arjun Berera.
\newblock Warm inflation.
\newblock {\em Physical Review Letters}, 75(18):3218–3221, October 1995.

\bibitem{Kamali_2023}
Vahid Kamali, Meysam Motaharfar, and Rudnei~O. Ramos.
\newblock Recent developments in warm inflation.
\newblock {\em Universe}, 9(3):124, February 2023.

\bibitem{Berera_2023}
Arjun Berera.
\newblock The warm inflation story.
\newblock {\em Universe}, 9(6):272, June 2023.

\bibitem{Trivedi:2020ljd}
Oem Trivedi.
\newblock {The exact solution approach to warm inflation}.
\newblock {\em Astropart. Phys.}, 158:102951, 2024.

\bibitem{Biswas_2024}
Sandip Biswas, Kaushik Bhattacharya, and Suratna Das.
\newblock Embedding ultraslow-roll inflaton dynamics in warm inflation.
\newblock {\em Physical Review D}, 109(2), January 2024.

\bibitem{Maartens_2010}
Roy Maartens and Kazuya Koyama.
\newblock Brane-world gravity.
\newblock {\em Living Reviews in Relativity}, 13(1), September 2010.

\bibitem{Langlois_2002}
David Langlois.
\newblock Brane cosmology.
\newblock {\em Progress of Theoretical Physics Supplement}, 148:181–212, 2002.

\bibitem{Brax_2004}
Philippe Brax, Carsten van~de Bruck, and Anne-Christine Davis.
\newblock Brane world cosmology.
\newblock {\em Reports on Progress in Physics}, 67(12):2183–2231, November 2004.

\bibitem{Deffayet2001}
C.~Deffayet.
\newblock Cosmology on a brane in minkowski bulk.
\newblock {\em Physics Letters B}, 502:199--208, 2001.

\bibitem{Sahni2003}
V.~Sahni and Y.~Shtanov.
\newblock Braneworld models of dark energy.
\newblock {\em Journal of Cosmology and Astroparticle Physics}, 2003(11):014, 2003.

\bibitem{Lue2006}
A.~Lue.
\newblock The phenomenology of dvali--gabadadze--porrati cosmologies.
\newblock {\em Physics Reports}, 423:1--48, 2006.

\bibitem{Maartens2006}
R.~Maartens and E.~Majerotto.
\newblock Observational constraints on the dgp braneworld model.
\newblock {\em Physical Review D}, 74:023004, 2006.

\bibitem{Herrera_2011}
Ramón Herrera and Eugenio San~Martin.
\newblock Warm-intermediate inflationary universe model in braneworld cosmologies.
\newblock {\em The European Physical Journal C}, 71(7), July 2011.

\bibitem{Cid_2007}
M~Antonella Cid, Sergio del Campo, and Ramón Herrera.
\newblock Warm inflation on the brane.
\newblock {\em Journal of Cosmology and Astroparticle Physics}, 2007(10):005–005, October 2007.

\bibitem{Panotopoulos_2007}
Grigoris Panotopoulos.
\newblock Assisted chaotic inflation in brane-world cosmology.
\newblock {\em Physical Review D}, 75(10), May 2007.

\bibitem{Trivedi:2021ivk}
Oem Trivedi.
\newblock {Rejuvenating the hope of a swampland consistent inflated multiverse with tachyonic inflation in the high-energy RS-II braneworld}.
\newblock {\em Mod. Phys. Lett. A}, 37(24):2250162, 2022.

\bibitem{BouhmadiLopez2011}
M.~Bouhmadi-L{\'o}pez, C.~Chen, and P.~Liu.
\newblock Inflation on the normal branch of dgp cosmology.
\newblock {\em Physical Review D}, 84:043505, 2011.

\bibitem{Hossain2016}
M.~Wali Hossain and Ratbay Myrzakulov.
\newblock Warm inflation in dgp braneworld cosmology.
\newblock {\em International Journal of Modern Physics D}, 25(16):1650091, 2016.

\bibitem{Germani_2002}
Cristiano Germani and Carlos~F. Sopuerta.
\newblock String inspired brane world cosmology.
\newblock {\em Physical Review Letters}, 88(23), May 2002.

\bibitem{doCarmo2016}
Manfredo~Perdig{\~a}o do~Carmo.
\newblock {\em Differential Geometry of Curves and Surfaces: Revised and Updated Second Edition}.
\newblock Dover Publications, Mineola, NY, 2016.

\bibitem{ONeill1983}
Barrett O'Neill.
\newblock {\em Semi-Riemannian Geometry With Applications to Relativity}.
\newblock Academic Press, New York, 1983.

\bibitem{Poisson2004}
Eric Poisson.
\newblock {\em A Relativist's Toolkit: The Mathematics of Black-Hole Mechanics}.
\newblock Cambridge University Press, Cambridge, 2004.

\bibitem{rs1}
Lisa Randall and Raman Sundrum.
\newblock {A Large mass hierarchy from a small extra dimension}.
\newblock {\em Phys. Rev. Lett.}, 83:3370--3373, 1999.

\bibitem{rs2}
Lisa Randall and Raman Sundrum.
\newblock {An Alternative to compactification}.
\newblock {\em Phys. Rev. Lett.}, 83:4690--4693, 1999.

\bibitem{Gogberashvili}
Merab Gogberashvili.
\newblock {Our world as an expanding shell}.
\newblock {\em EPL}, 49:396--399, 2000.

\bibitem{dgp}
G.~R. Dvali, Gregory Gabadadze, and Massimo Porrati.
\newblock {4-D gravity on a brane in 5-D Minkowski space}.
\newblock {\em Phys. Lett. B}, 485:208--214, 2000.

\bibitem{Dimopoulos_2017}
Konstantinos Dimopoulos.
\newblock Ultra slow-roll inflation demystified.
\newblock {\em Physics Letters B}, 775:262–265, December 2017.

\bibitem{Martin_2013}
Jérôme Martin, Hayato Motohashi, and Teruaki Suyama.
\newblock Ultra slow-roll inflation and the non-gaussianity consistency relation.
\newblock {\em Physical Review D}, 87(2), January 2013.

\bibitem{Kinney_2005}
William~H. Kinney.
\newblock Horizon crossing and inflation with large $\eta$.
\newblock {\em Physical Review D}, 72(2), July 2005.

\bibitem{Di_Marco_2024}
Alessandro Di~Marco, Emanuele Orazi, and Gianfranco Pradisi.
\newblock Introduction to the number of e-folds in slow-roll inflation.
\newblock {\em Universe}, 10(7):284, June 2024.

\bibitem{PhysRevD.105.083525}
Sina Hooshangi, Alireza Talebian, Mohammad~Hossein Namjoo, and Hassan Firouzjahi.
\newblock Multiple field ultraslow-roll inflation: Primordial black holes from straight bulk and distorted boundary.
\newblock {\em Phys. Rev. D}, 105:083525, Apr 2022.

\bibitem{BASTERO_GIL_2009}
Mat Basteto-Gil and Arjun Berera.
\newblock Warm inflation model building.
\newblock {\em International Journal of Modern Physics A}, 24(12):2207–2240, May 2009.

\bibitem{2023Univ....9..124K}
Vahid {Kamali}, Meysam {Motaharfar}, and Rudnei~O. {Ramos}.
\newblock {Recent Developments in Warm Inflation}.
\newblock {\em Universe}, 9(3):124, February 2023.

\bibitem{Motaharfar_2019}
Meysam Motaharfar, Vahid Kamali, and Rudnei~O. Ramos.
\newblock Warm inflation as a way out of the swampland.
\newblock {\em Physical Review D}, 99(6), March 2019.

\bibitem{Das_2020}
Suratna Das and Rudnei~O. Ramos.
\newblock Runaway potentials in warm inflation satisfying the swampland conjectures.
\newblock {\em Physical Review D}, 102(10), November 2020.

\bibitem{Kamali_2020}
Vahid Kamali, Meysam Motaharfar, and Rudnei~O. Ramos.
\newblock Warm brane inflation with an exponential potential: A consistent realization away from the swampland.
\newblock {\em Physical Review D}, 101(2), January 2020.

\bibitem{Raveri:2018ddi}
Marco Raveri, Wayne Hu, and Savdeep Sethi.
\newblock Swampland conjectures and late-time cosmology.
\newblock {\em Phys. Rev. D}, 99(8):083518, 2019.

\bibitem{Hamaguchi:2018vtv}
Koichi Hamaguchi, Masahiro Ibe, and Takeo Moroi.
\newblock The swampland conjecture and the higgs expectation value.
\newblock {\em JHEP}, 12:023, 2018.

\bibitem{Fukuda:2018haz}
Hajime Fukuda, Ryo Saito, Satoshi Shirai, and Masahito Yamazaki.
\newblock Phenomenological consequences of the refined swampland conjecture.
\newblock {\em Phys. Rev. D}, 99(8):083520, 2019.

\bibitem{Trivedi:2020wxf}
Oem Trivedi.
\newblock {Swampland conjectures and single-field inflation in nonstandard cosmological scenarios}.
\newblock {\em Int. J. Mod. Phys. D}, 32(01):2250130, 2023.

\bibitem{GargKrishnan2019BoundsSlowRollSwampland}
Sumit~K. Garg and Chethan Krishnan.
\newblock Bounds on slow roll and the de sitter swampland.
\newblock {\em Journal of High Energy Physics}, 2019(11):075, 2019.
\newblock arXiv:1807.05193 [hep‐th].

\bibitem{Benetti_2017}
Micol Benetti and Rudnei~O. Ramos.
\newblock Warm inflation dissipative effects: Predictions and constraints from the planck data.
\newblock {\em Physical Review D}, 95(2), January 2017.

\bibitem{Bastero_Gil_2016}
Mar Bastero-Gil, Arjun Berera, Rudnei~O. Ramos, and João~G. Rosa.
\newblock Warm little inflaton.
\newblock {\em Physical Review Letters}, 117(15), October 2016.

\bibitem{Bastero_Gil_2013}
Mar Bastero-Gil, Arjun Berera, Rudnei~O Ramos, and João~G Rosa.
\newblock General dissipation coefficient in low-temperature warm inflation.
\newblock {\em Journal of Cosmology and Astroparticle Physics}, 2013(01):016–016, January 2013.

\bibitem{kamali2016tachyonwarmintermediateinflationlight}
Vahid Kamali, Spyros Basilakos, and Ahmad Mehrabi.
\newblock Tachyon warm-intermediate inflation in the light of planck data, 2016.

\bibitem{Sasankan_2017}
N.~Sasankan, Mayukh~R. Gangopadhyay, G.~J. Mathews, and M.~Kusakabe.
\newblock Limits on brane-world and particle dark radiation from big bang nucleosynthesis and the cmb.
\newblock {\em International Journal of Modern Physics E}, 26(08):1741007, August 2017.

\bibitem{Bastero_Gil_2019}
Mar Bastero-Gil, Arjun Berera, and Jaime~R. Calderón.
\newblock Reexamination of the warm inflation curvature perturbations spectrum.
\newblock {\em Journal of Cosmology and Astroparticle Physics}, 2019(07):019–019, July 2019.

\bibitem{Hall2004}
L.~M.~H. Hall, I.~G. Moss, and A.~Berera.
\newblock Scalar perturbation spectra from warm inflation.
\newblock {\em Physical Review D}, 69:083525, 2004.

\bibitem{Moss2008}
I.~G. Moss and C.~Xiong.
\newblock Dissipation coefficients for supersymmetric inflatonary models.
\newblock {\em Journal of Cosmology and Astroparticle Physics}, 2008(11):023, 2008.

\bibitem{BasteroGil2011}
M.~Bastero-Gil, A.~Berera, and R.~O. Ramos.
\newblock Shear viscous effects on the primordial power spectrum from warm inflation.
\newblock {\em Journal of Cosmology and Astroparticle Physics}, 2011(07):030, 2011.

\bibitem{Graham2009}
C.~Graham and I.~G. Moss.
\newblock Density fluctuations from warm inflation.
\newblock {\em Journal of Cosmology and Astroparticle Physics}, 2009(07):013, 2009.

\bibitem{Byrnes2012_PBH_NG_Constraint}
Christian~T. Byrnes, Edmund~J. Copeland, and Anne~M. Green.
\newblock Primordial black holes as a tool for constraining non‐gaussianity.
\newblock {\em Phys. Rev. D}, 86(4):043512, 2012.

\bibitem{YoungMuscoByrnes2019_NonlinearRelation}
Sam Young, Ilia Musco, and Christian~T. Byrnes.
\newblock Primordial black hole formation and abundance: contribution from the non‐linear relation between the density and curvature perturbation.
\newblock {\em JCAP}, 11:012, 2019.

\bibitem{KehagiasMuscoRiotto2019_Threshold}
Alex Kehagias, Ilia Musco, and Antonio Riotto.
\newblock Non‐gaussian formation of primordial black holes: Effects on the threshold.
\newblock {\em JCAP}, 12:029, 2019.

\bibitem{DeLuca2019_IneludibleNG}
Valerio~De Luca, Gabriele Franciolini, Alex Kehagias, Massimo Peloso, Antonio Riotto, and Caner Ünal.
\newblock The ineludible non‐gaussianity of the primordial black hole abundance.
\newblock {\em JCAP}, (07):048, 2019.

\bibitem{Gow2022_NonPerturbative}
Andrew~D. Gow, Hooshyar Assadullahi, Joseph H.~P. Jackson, Kazuya Koyama, Vincent Vennin, and David Wands.
\newblock Non‐perturbative non‐gaussianity and primordial black holes.
\newblock {\em JCAP}, 2023.

\bibitem{Pi2024_ReviewNG_PBH}
Shi Pi.
\newblock Non‐gaussianities in primordial black hole formation and induced gravitational waves.
\newblock {\em arXiv preprint arXiv:2404.06151}, 2024.
\newblock Comprehensive review.

\bibitem{Pi_2025}
Shi Pi.
\newblock {\em Non-Gaussianities and Primordial Black Holes}, page 155–200.
\newblock Springer Nature Singapore, 2025.

\bibitem{Montefalcone2023}
G.~Montefalcone, R.~O. Ramos, and A.~Berera.
\newblock Non-gaussianity in warm inflation revisited.
\newblock {\em Physical Review D}, 107:083520, 2023.

\bibitem{Arya2019}
R.~Arya and S.~Das.
\newblock Primordial black holes from warm inflation and their detection.
\newblock {\em Journal of Cosmology and Astroparticle Physics}, 2019(09):034, 2019.

\end{thebibliography}
\end{document}